\documentclass[]{emulateapj}
\usepackage{natbib,amsmath}

\begin{document}

\def \ngdz  {1121}   
\def \ngal  {66}     
\def \nmiss  {48}    
\def \npcand {1450}  
\def \nprim  {1040}  
\def \newagn {217}   
\def \agnrate {$86\%$} 
\def \mqv  {v4.5}    
\def \mzem {z_{\rm em}}
\def \zem {$\mzem$}
\def \zbg {$z_{\rm bg}$}
\def \zfg {$z_{\rm fg}$}
\def \mzfg {z_{\rm fg}}
\def \zlya {$z_{\rm Ly\alpha}$}
\def \wlya {$W_{\rm Ly\alpha}$}
\def \mwlya {W_{\rm Ly\alpha}}
\def \lbol {$L_{\rm Bol}$}
\def \guv {$g_{\rm UV}$}
\def \nhi  {$N_{\rm HI}$}
\def \mnhi  {N_{\rm HI}}
\def \kms  {\, km~s$^{-1}$}
\def \mkms  {{\rm km~s^{-1}}}
\def \lya  {Ly$\alpha$}
\def \mlya  {{\rm Ly\alpha}}
\def \lyb  {Ly$\beta$}
\def \hMpc      {h^{-1}{\rm\ Mpc}}
\def \msol      {{\rm\ M}_\odot}
\def\cm#1{\, {\rm cm^{#1}}}
\def \cgsflux   {{\rm erg\ s^{-1}\ cm^{-2}}}
\def \cgssflux   {{\rm erg\ s^{-1}\ Hz^{-1} cm^{-2}}}
\def\sci#1{{\; \times \; 10^{#1}}}
\def\N#1{{N({\rm #1})}}
\def \pasa {PASA}

\title{The UV-bright Quasar Survey (UVQS):  DR1}

\author{
TalaWanda R. Monroe\altaffilmark{1},
J. Xavier Prochaska\altaffilmark{2},
N. Tejos\altaffilmark{2},
G. Worseck\altaffilmark{3},
Joseph F. Hennawi\altaffilmark{3},
Tobias Schmidt\altaffilmark{3},
Jason Tumlinson\altaffilmark{1},
Yue Shen\altaffilmark{4}
}
\altaffiltext{1}{Space Telescope Science Institute, Baltimore, MD, 21218, USA}
\altaffiltext{2}{Department of Astronomy and Astrophysics, UCO/Lick
  Observatory, University of California, 1156 High Street, Santa Cruz,
  CA 95064, USA}
\altaffiltext{3}{Max-Planck-Institut f\"ur Astronomie, K\"onigstuhl 17,
  D-69115 Heidelberg, Germany} 
\altaffiltext{4}{Department of Astronomy and National Center for Supercomputing Applications, University of Illinois at Urbana-Champaign, Urbana, IL 61801, USA }

\begin{abstract}
We present the first data release (DR1) from our
UV-bright Quasar Survey (UVQS) for new $z \sim 1$
active galactic nuclei (AGN) across the sky.  Using simple GALEX UV and
WISE near-IR color selection criteria, we generated a list of 
\npcand\ primary candidates with $FUV < 18.5$\,mag. 
We obtained discovery
spectra, primarily on 3m-class telescopes, for \nprim\ 
of these candidates and confirmed 86\%\ as AGN with
redshifts generally at $z>0.5$.
Including a small set of observed secondary candidates,
we report the discovery of \newagn\ AGN with $FUV < 18$\,mag 
that had no previously reported spectroscopic redshift.
These are excellent potential targets for UV spectroscopy
before the end of the {\it Hubble Space Telescope}
mission.  The main data products are publicly released
through the Mikulski Archive for Space Telescopes.

\end{abstract}


\section{Introduction}

Presently, the only efficient means of studying the diffuse gas
surrounding galaxies (a.k.a. halo gas or the circumgalactic medium,
CGM) and in between galaxies (a.k.a. the intergalactic medium, IGM) is
through absorption-line spectroscopy of luminous, background quasars
\citep[e.g.][]{tripp08,tumlinson+13,tejos+14}.  
Furthermore, because the principal
transitions to diagnose gas lie at far-ultraviolet
(FUV) wavelengths ($\lambda_{\rm rest} < 2000$\AA), 
for $z<1$ studies one requires UV spectrometers on space-borne
facilities.   Currently, and for the foreseeable future, 
the Hubble Space Telescope ({\it HST}) affords the only opportunity
for such research, primarily with the Cosmic Origins Spectrograph (COS).
Given the modest aperture of {\it HST},
these observations
are generally restricted to the brightest FUV
quasars on the sky.

High-quality, FUV spectroscopy of $z \sim 1$ quasars have enabled
several, unique experiments to study the CGM and IGM of the universe
over the past $\sim 10$\,Gyr.  These include:
 (1) the survey of highly ionized gas via the NeVIII~$\lambda\lambda
 770,780$ doublet and/or broad \ion{H}{1} \lya\ systems
 that may trace the elusive warm-hot ionized medium
 \citep[WHIM; e.g.][]{lsr+07,mtw+13,tejos+15}; 
 (2) the search for signatures of galactic and AGN feedback
 \citep[e.g.][]{tmp+11}; 
 (3) the measurements of enrichment in galactic halos and 
 optically thick gas \citep[e.g.][]{lht+13,werk+13,werk+14};
 and
 (4) revealing the structure of the cosmic web and its correlation to
 the large-scale structures traced by galaxies \citep[e.g.][]{tejos+14}.
While each of these programs has had scientific impact, 
they are limited by sample variance.

An efficient way to increase the volumes surveyed is to focus 
on those bright UV QSOs 
that maximize the redshift path covered, i.e. those
with $z_{\rm em} \gtrsim 1$.
To date, only a small number of $z \sim 1$
quasars have been observed with {\it HST}, primarily 
corresponding to the set of sources with
very high FUV flux. These have been
drawn from historical, large-area surveys for AGN 
(e.g.\ the Palomar-Green Bright Quasar Survey and the
Hamburg/ESO survey)
and more recently the Northern Galactic pole footprint of the Sloan Digital Sky Survey (SDSS). 
Cross-matching the quasar sample of \cite{milliq} against the 
point-source catalog of the GALEX survey, one recovers $\approx 140$ sources
with $z>0.6$ and FUV~$<18$\,mag (fewer than 50 at $z>1$). 
These are preferentially located within the SDSS footprint 
which has extensively surveyed
the Northern galactic pole for quasars \citep[e.g.][]{sdss_qso_dr7}.
Given that {\it HST} may observe nearly any position on the sky, we 
are motivated to perform an all-sky search for new, FUV-bright
quasars across the sky.  
Indeed, progress in this area demands the discovery
of new FUV-bright quasars.

The principal goal of our survey is to provide the community
with a nearly complete set of UV-bright, AGN
before the termination of the {\it HST} mission.
We recognized that the combination of two NASA imaging missions --
GALEX and WISE -- enables a modern, all-sky search for 
UV bright quasars.  
These must be spectroscopically confirmed, however,
before subsequent {\it HST} observations.
Given our interest in FUV-bright
sources, this implies optically bright candidates that can be 
spectroscopically confirmed on 3m-class telescopes.
The following manuscript provides the first data release
(DR1) from our UV-bright Quasar Survey (UVQS).
The main data products are available at the
Mikulski Archive for Space Telescopes\footnote{https://archive.stsci.edu/prepds/uvqs}.

This paper is organized as follows.  Section~\ref{sec:sample}
describes the UVQS candidate selection, focused on detecting
$z \sim 1$ quasars with FUV~$< 18$\,mag.
The follow-up spectroscopy is discussed in Section~\ref{sec:obs}
and the redshift analysis is described in Section~\ref{sec:z}.
We present the primary results in Section~\ref{sec:results}.
When relevant, we assume a $\Lambda$CDM cosmology with $h=0.7,
\Omega_m=0.3,$ and $\Omega_\Lambda=0.7$.

\section{The UVQS Candidates}
\label{sec:sample}

\begin{figure}
\includegraphics[width=3.5in]{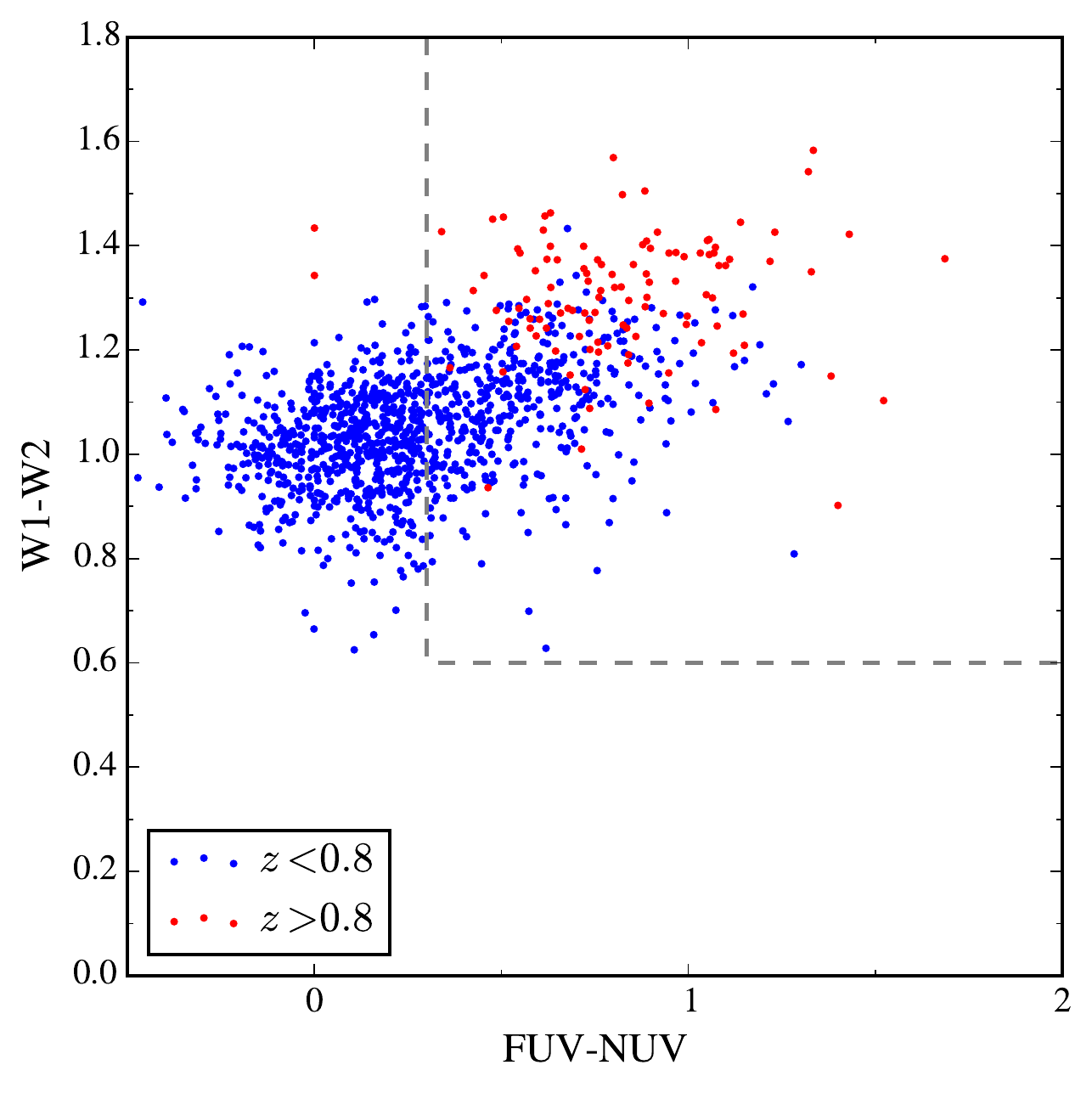}
\caption{Color-color plot of WISE and GALEX photometry of the SDSS
DR7 quasars \citep{sdss_qso_dr7} that have a NUV flux $<19$\,mag.
It is evident that each has a $W1-W2>0.6$\,mag color, consistent
with the \cite{stern12} selection-criteria for AGN.
Furthermore, the $z>0.8$ quasars exhibit redder $FUV-NUV$ colors
which we hypothesize results from intervening Lyman limit opacity.
The gray dashed lines at $W1-W2 = 0.6$\,mag and
$FUV-NUV=0.3$\,mag indicate the color-color criteria adopted
for our primary candidates (Table~\ref{tab:candidates}).
}
\label{fig:color_cuts}
\end{figure}

With the explicit goal of discovering new FUV-bright quasars
at $z \sim 1$ across the sky, we developed color-color criteria 
leveraging the all-sky surveys of the WISE and GALEX missions to:
 (i) isolate AGN; and
 (ii) maximize the probability that these AGN
 lay at $\mzem \gtrsim 1$.
For the first criterion, we followed the impressive results from the
WISE team who demonstrated the clean separation of AGN from stars,
galaxies and other astrophysical sources using WISE photometry
\citep{stern12}. 
Specifically, \cite{stern12} showed that AGN tend to exhibit 
$W1-W2 > 0.4$\,mag with galaxies and stars having smaller values.
Although this criterion may not capture all AGN 
\citep[e.g.][]{assef+10}, we strongly expect
that every UV-bright AGN satisfies the criterion.
Indeed, we find that of the 1148 quasars at $z<1.5$
from SDSS DR7 detected by GALEX ($NUV < 19.0$), all have
$W1-W2 > 0.625$\,mag (Figure~\ref{fig:color_cuts}).
The overwhelming majority of these have $z<0.8$ (90\%).

Figure~\ref{fig:color_cuts} also shows the $FUV-NUV$ colors
of these quasars.  These were measured from the ``photoobjall"
catalog of the GALEXGR6Plus7 context at MAST and improved,
where possible, using the MIS catalog 
\citep[``bcscat\_mis"][]{bianchi+14}.  
We see that the majority of $z<0.8$ quasars
have $FUV-NUV < 0.3$\,mag (60\%) and that nearly all of the 
$z>0.8$ quasars have a redder $FUV-NUV$ color.
We believe that this `reddening' primarily results from the presence of
one or more Lyman limit systems (LLSs) in the redshift interval
$0.5 < z < 0.8$ whose continuum opacity reduces only the FUV flux. 
We infer that nearly every $z \sim 1$ quasar exhibits at least one
intervening LLS\footnote{In standard IGM nomenclature, LLS with
$\mnhi < 10^{17.3} \cm{-2}$ are often referred to as partial LLS or pLLS.	
} with $\mnhi > 10^{16.7} \cm{-2}$.

\begin{figure*}
\begin{center}
\includegraphics[width=6in]{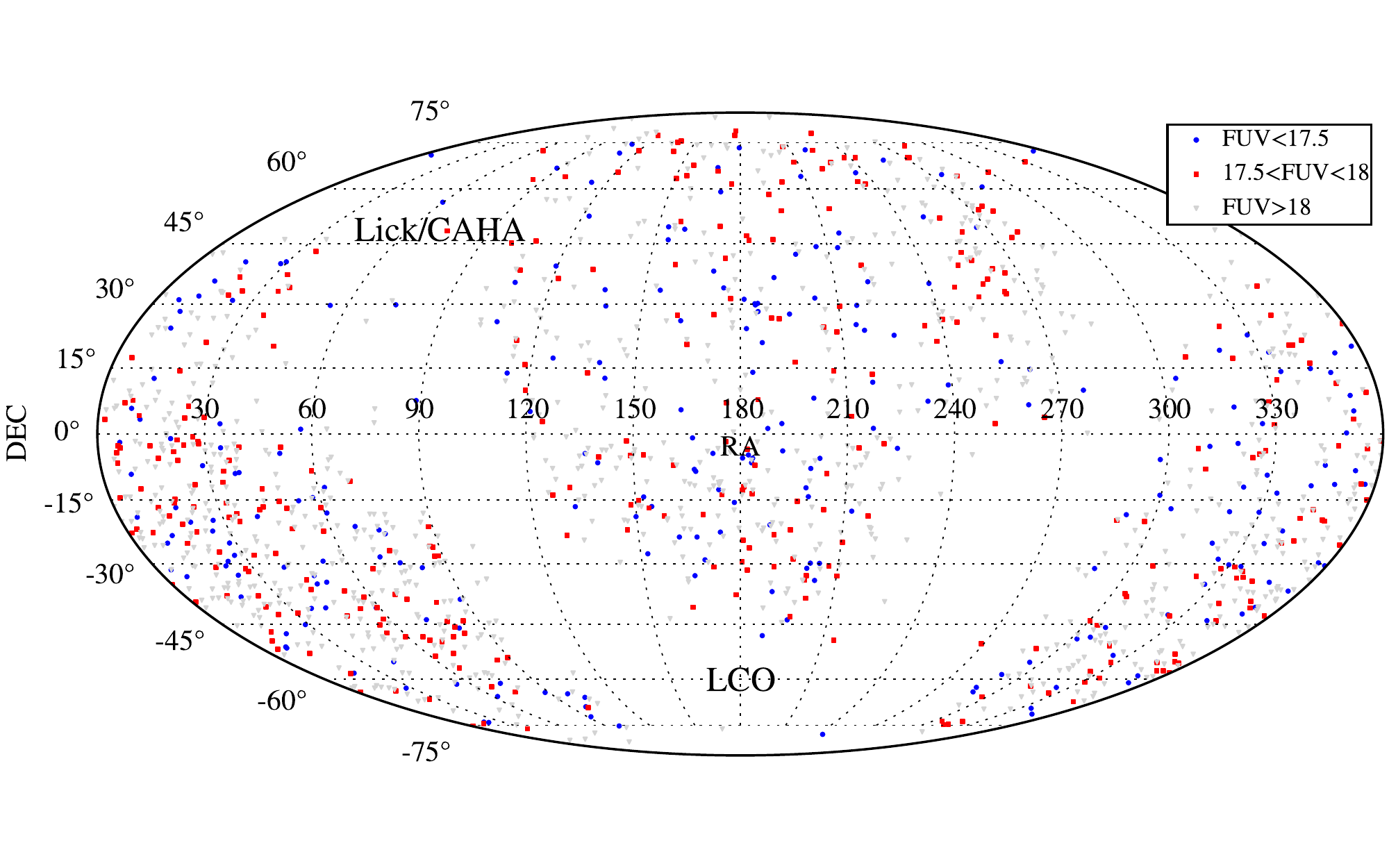}
\caption{An all-sky plot describing the spatial distribution of
our primary candidates, coded by FUV flux.  We have avoided the
Galactic plane and one also notes fewer targets towards the
Northern Galactic pole (i.e.\ within the SDSS footprint).
}
\label{fig:candidates}
\end{center}
\end{figure*}

With our photometric criteria established,

\begin{align} 
W1-W2 > 0.6\,{\rm mag} \\
FUV-NUV > 0.3\,{\rm mag} \\
FUV < 18.5\,{\rm mag},
\end{align} 
we cross-matched every source in the GALEXGR6Plus7 catalogs\footnote{
Our explicit cassjobs query for the AIS data was:	
select objid, ra, dec, fuv\_mag as fuv, nuv\_mag as nuv from photoobjall;
where fuv\_mag BETWEEN 12. and 18.5;
and (fuv\_mag-nuv\_mag) BETWEEN -0.5 and 2.0;
and fuv\_mag $> -999$;
and nuv\_mag $> -999$.
We then used the following for the MIS to improve the photometry: 
select objid, ra, dec, fuv\_mag as fuv, nuv\_mag as nuv from bcscat\_mis;
where fuv\_mag BETWEEN 12. and 18.5;
and (fuv\_mag-nuv\_mag) BETWEEN -0.5 and 2.0;
and fuv\_mag $> -999$;
and nuv\_mag $> -999$.
}
satisfying these criteria against the AllWISE Source Catalog.
To avoid selecting already known quasars given the beam sizes of WISE and GALEX, we then eliminated any sources that lay within $5''$ of a
UV-bright quasar from SDSS DR7. 
This generated a list of \npcand\ primary candidates 
(Table~\ref{tab:candidates}).
We discovered, during our analysis, that this 
candidate list includes hundreds
of previously cataloged sources from other surveys.
This includes the SDSS-III survey which included  
WISE-selected quasar targets (Paris et al.\ 2015). 
Their primary WISE criteria, however, precluded overlap
with our sample.
Given that several of these surveys have known examples of
false redshift identifications or
do not provide the discovery spectra, we maintained the list and
re-observed many of the brighter sources ($FUV < 18$\,mag).
Figure~\ref{fig:candidates} shows an all-sky summary of the
UVQS candidates, separated by $FUV$ flux. 
The exclusion of the
Galactic plane is obvious and the lower
incidence of sources in the SDSS footprint is notable.

\begin{deluxetable*}{lrrcccc}
\tablewidth{0pc}
\tablecaption{UVQS DR1 PRIMARY CANDIDATES\label{tab:candidates}}
\tabletypesize{\scriptsize}
\tablehead{\colhead{Name} & \colhead{$\alpha_{\rm J2000}$} 
& \colhead{$\delta_{\rm J2000}$} & \colhead{W1} & \colhead{W2} 
& \colhead{FUV} & \colhead{NUV} \\ 
& (deg) & (deg) & (mag) & (mag) & (mag) & (mag) } 
\startdata 
UVQSJ000000.15$-200427.7$&0.00064 & $-20.07437$&13.55 & 12.54&18.27 & 17.97\\ 
UVQSJ000002.92$-350332.6$&0.01218 & $-35.05905$&12.69 & 11.55&17.61 & 17.31\\ 
UVQSJ000009.66$-163441.5$&0.04023 & $-16.57819$&13.43 & 12.19&18.48 & 17.72\\ 
UVQSJ000037.52$-752442.6$&0.15633 & $-75.41184$&11.69 & 10.63&17.81 & 17.45\\ 
UVQSJ000355.89$-224122.4$&0.98286 & $-22.68955$&13.24 & 12.11&17.97 & 17.24\\ 
UVQSJ000503.70$-391747.9$&1.26542 & $-39.29664$&12.26 & 11.12&17.82 & 17.23\\ 
UVQSJ000609.57$-261140.6$&1.53989 & $-26.19460$&13.31 & 12.12&18.16 & 17.53\\ 
UVQSJ000613.29$+321534.6$&1.55537 & $32.25960$&12.93 & 11.75&18.42 & 17.95\\ 
UVQSJ000717.70$+421646.7$&1.82374 & $42.27963$&12.44 & 11.51&18.09 & 17.61\\ 
UVQSJ000741.01$-635145.9$&1.92085 & $-63.86274$&12.65 & 11.45&17.96 & 17.41\\ 
UVQSJ000750.79$+031733.1$&1.96161 & $3.29253$&12.98 & 11.58&17.80 & 17.01\\ 
UVQSJ000755.68$+052818.8$&1.98200 & $5.47189$&13.12 & 11.73&18.07 & 17.29\\ 
UVQSJ001444.03$-223522.6$&3.68344 & $-22.58961$&13.16 & 11.77&18.39 & 17.34\\ 
\enddata 
\tablecomments{Table \ref{tab:candidates} is published in its entirety in the electronic edition, a portion is shown here for guidance regarding its form and content.}
\end{deluxetable*}

\begin{deluxetable*}{lrrcccc}
\tablewidth{0pc}
\tablecaption{UVQS DR1 SECONDARY CANDIDATES\label{tab:sec}}
\tabletypesize{\scriptsize}
\tablehead{\colhead{Name} & \colhead{$\alpha_{\rm J2000}$} 
& \colhead{$\delta_{\rm J2000}$} & \colhead{W1} & \colhead{W2} 
& \colhead{FUV} & \colhead{NUV} \\ 
& (deg) & (deg) & (mag) & (mag) & (mag) & (mag) } 
\startdata 
UVQSJ000007.85$-633535.2$&0.03271 & $-63.59311$&13.25 & 12.32&18.06 & 17.77\\ 
UVQSJ000011.73$+052317.4$&0.04886 & $5.38818$&11.90 & 10.88&18.37 & 18.30\\ 
UVQSJ000024.03$-275153.5$&0.10013 & $-27.86486$&12.85 & 11.80&18.32 & 18.14\\ 
UVQSJ000024.42$-124547.9$&0.10173 & $-12.76331$&11.08 & 10.08&15.82 & 15.78\\ 
UVQSJ000036.68$-634123.7$&0.15285 & $-63.68991$&12.44 & 11.46&18.10 & 18.15\\ 
UVQSJ000053.51$-443933.5$&0.22297 & $-44.65930$&12.56 & 11.81&17.95 & 17.95\\ 
UVQSJ000054.29$+183021.4$&0.22621 & $18.50594$&13.26 & 12.18&16.65 & 16.47\\ 
UVQSJ000055.97$+172338.9$&0.23320 & $17.39414$&13.13 & 12.09&17.71 & 17.83\\ 
UVQSJ000103.53$-114725.9$&0.26469 & $-11.79053$&12.70 & 11.59&18.04 & 18.13\\ 
UVQSJ000115.89$+051902.1$&0.31621 & $5.31725$&13.47 & 12.61&18.43 & 18.44\\ 
UVQSJ000118.99$+172425.3$&0.32913 & $17.40703$&12.86 & 11.88&18.48 & 18.33\\ 
UVQSJ000128.58$-320842.1$&0.36908 & $-32.14502$&13.17 & 12.05&18.30 & 18.03\\ 
UVQSJ000146.09$-765714.3$&0.44203 & $-76.95396$&11.01 & 10.23&17.05 & 16.88\\ 
UVQSJ000150.56$+111647.3$&0.46068 & $11.27981$&11.68 & 10.73&17.27 & 17.12\\ 
UVQSJ000200.53$-073907.5$&0.50220 & $-7.65209$&14.11 & 13.01&18.19 & 18.13\\ 
UVQSJ000210.06$+171558.2$&0.54193 & $17.26616$&15.50 & 14.85&18.46 & 18.16\\ 
\enddata 
\tablecomments{Table \ref{tab:sec} is published in its entirety in the electronic edition, a portion is shown here for guidance regarding its form and content.}
\end{deluxetable*}

In several of the observing runs, conditions were unexpectedly
favorable and we exhausted the primary candidates at certain RA
ranges.  To fill the remaining observing time, we generated a 
secondary candidate list with one criterion modified:
 $-0.5 < FUV-NUV < 0.3$.  This would permit a much
higher fraction of
low-$z$ AGN, but may also yield a few sources at $z \sim 1$.  
This secondary set of candidates is provided in 
Table~\ref{tab:sec}.

\begin{figure}[ht]
\includegraphics[width=3.5in]{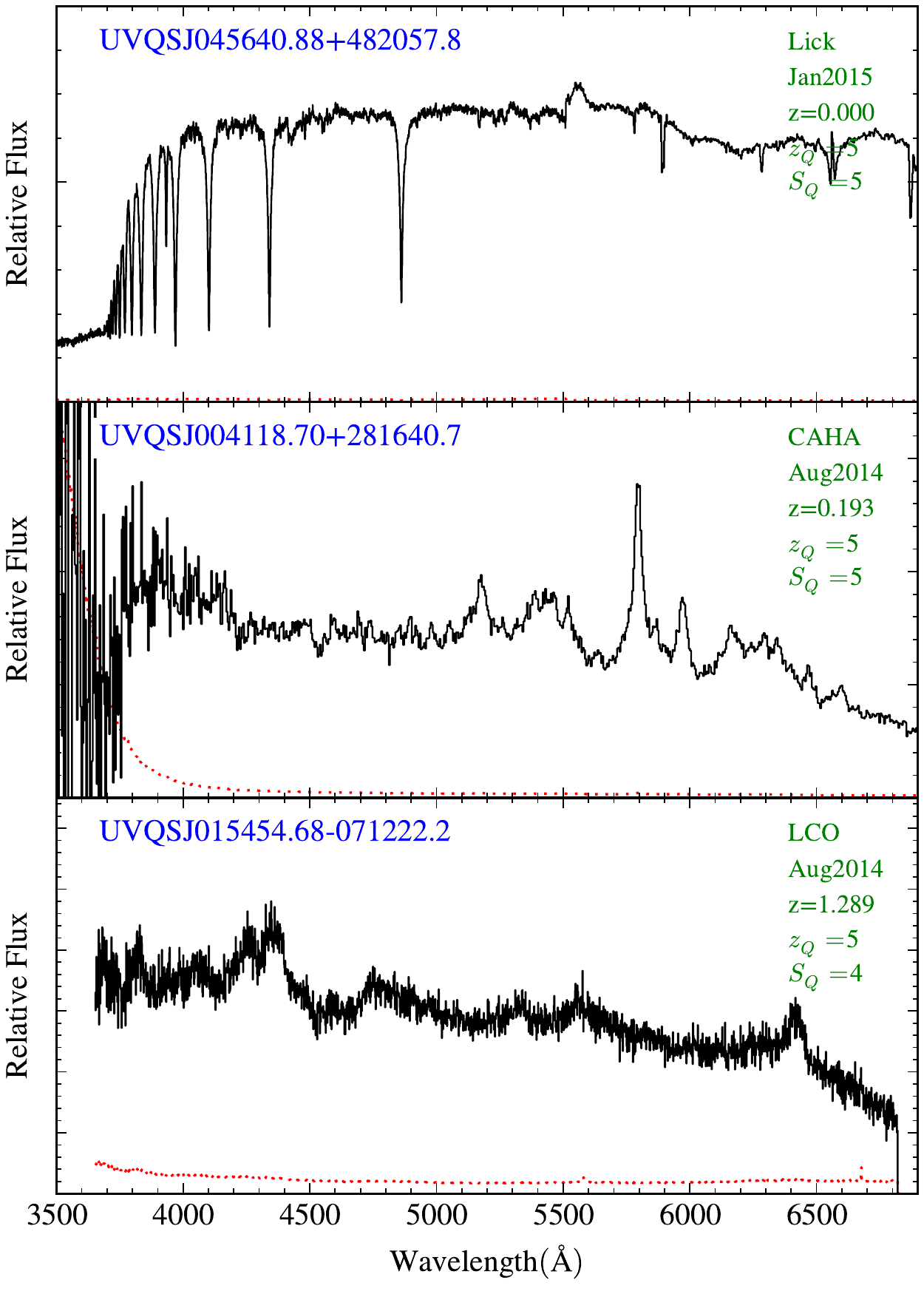}
\caption{Characteristic spectra of the UVQS DR1 data release.
From top to bottom, we show examples of a Galactic star,
a low-$z$ AGN, and a $z>1$ quasar.
}
\label{fig:ex_spec}
\end{figure}

\section{Observations and Data Processing}
\label{sec:obs}

We proceeded to obtain discovery-quality longslit spectra
(i.e.\ low-dispersion, large wavelength coverage,
modest signal-to-noise (S/N) of our UVQS candidates in one calendar year.
Our principal facilities were:
 (i) the dual Kast spectrometer on the 3m Shane telescope at Lick Observatory;
 (ii) the Boller \& Chivens (BCS) spectrometer on 
 the Ir\'en\'ee du Pont $100"$ telescope at Las Campanas Observatory;
 and
 (iii) the Calar Alto Faint Object Spectrograph (CAFOS) 
 on the CAHA 2.2m telescope at Calar Alto Observatory (CAHA).
We acquired an additional $\approx 20$~spectra 
on larger aperture telescopes
(Keck/ESI, MMT/MBC, Magellan/MagE) 
during twilight or under poor observing conditions.
Typical exposure times were limited to $\lesssim 200$\,s with adjustments
for fainter sources or sub-optimal observing conditions.
Table~\ref{tab:journal} provides a list of the 
observed candidates.

The two-dimensional (2D) spectral images and calibration frames were
reduced with custom software, primarily the LowRedux 
package\footnote{http://www.ucolick.org/$\sim$xavier/LowRedux/} 
developed by J. Hennawi, X. Prochaska, and D. Schlegel.
Briefly, the images were bias subtracted, flat-fielded using quartz
lamp spectral images, and wavelength calibrated with arc lamp
exposures.  Objects within the slit were automatically identified
and optimally extracted to generate 1D spectra.
These were fluxed after generating a sensitivity function
from observations of spectrophotometric standard stars taken
during each observing run.  We did not carefully account for
varying atmospheric conditions nor did
we correct for slit-losses
from variable seeing or atmospheric dispersion.
Therefore, the reported fluxes are crude and not even 
especially accurate in a
relative sense, particularly at the wavelength extrema.
Although we occasionally obtained multiple exposures for a given
source, these were not combined;
the highest quality spectrum was analyzed. 
Upon visual inspection we assigned a spectral data quality number 
(SPEC\_QUAL) to each spectrum.  
Our scale spans 0 to 5, in which 0 is poor, or unusable, and 5 is excellent.  
SPEC\_QUAL values are a good proxy for S/N ratio
and are included in Table~\ref{tab:journal}.
Note, even spectra without spectral features may have a 
high SPEC\_QUAL value.

\begin{deluxetable}{lcccc}
\tablewidth{0pc}
\tablecaption{UVQS DR1 OBSERVATIONS\label{tab:journal}}
\tabletypesize{\scriptsize}
\tablehead{\colhead{Name} & \colhead{Obs.} & \colhead{Inst.} 
& \colhead{Date} & \colhead{Q$^a$} } 
\startdata 
UVQSJ000000.15$-200427.7$&LCO&BCS&Aug2014&3\\ 
UVQSJ000009.65$-163441.4$&LCO&BCS&Aug2014&3\\ 
UVQSJ000503.70$-391747.9$&LCO&BCS&Aug2014&3\\ 
UVQSJ000609.57$-261140.5$&LCO&BCS&Aug2014&3\\ 
UVQSJ000613.28$+321534.5$&Lick&Kast&Jan2015&2\\ 
UVQSJ000717.69$+421646.6$&Lick&Kast&Jan2015&4\\ 
UVQSJ000741.00$-635145.8$&LCO&BCS&Aug2014&3\\ 
UVQSJ000750.78$+031733.1$&LCO&BCS&Aug2014&4\\ 
UVQSJ000755.67$+052818.8$&LCO&BCS&Aug2014&3\\ 
UVQSJ000814.35$+121201.3$&Lick&Kast&Jan2015&1\\ 
UVQSJ000856.77$-235317.5$&LCO&BCS&Aug2014&4\\ 
UVQSJ001015.62$-624045.1$&LCO&BCS&Aug2014&3\\ 
UVQSJ001121.73$-200212.1$&LCO&BCS&Aug2014&3\\ 
UVQSJ001155.60$-240438.8$&LCO&BCS&Aug2014&4\\ 
UVQSJ001444.02$-223522.6$&LCO&BCS&Aug2014&3\\ 
UVQSJ001521.62$-385419.1$&LCO&BCS&Aug2014&3\\ 
UVQSJ001529.53$-360535.3$&LCO&BCS&Aug2014&3\\ 
UVQSJ001637.90$-054424.8$&Lick&Kast&Jan2015&3\\ 
UVQSJ001641.88$-312656.6$&Magellan&MagE&Jul2014&5\\ 
UVQSJ001653.66$-530932.6$&LCO&BCS&Aug2014&3\\ 
UVQSJ001655.68$+054822.9$&LCO&BCS&Aug2014&3\\ 
\enddata 
\tablenotetext{a}{Spectral quality:  0=Too poor for analysis; 5=Excellent} 
\tablecomments{Table \ref{tab:journal} is published in its entirety in the electronic edition, a portion is shown here for guidance regarding its form and content.}
\end{deluxetable}

The calibrated 1D spectra are published in DR1 and provided
at https://archive.stsci.edu/prepds/uvqs.
We also present a cutout, optical image of each source taken
from the SDSS or DSS survey.
Figure~\ref{fig:ex_spec} shows representative spectra from
the UVQ DR1 sample, including examples of a Galactic star,
a low-$z$ AGN, and a $z>1$
quasar (PHL~1288).  At the S/N of these spectra (each of which
has a spectral quality of 4 or 5), redshift identification is
straightforward.  We note that $\approx 50\%$ of our spectra
have this data quality and another $40\%$ have SPEC\_QUAL=3,
which we consider sufficient for redshift analysis.

\begin{figure}
\includegraphics[width=3.5in]{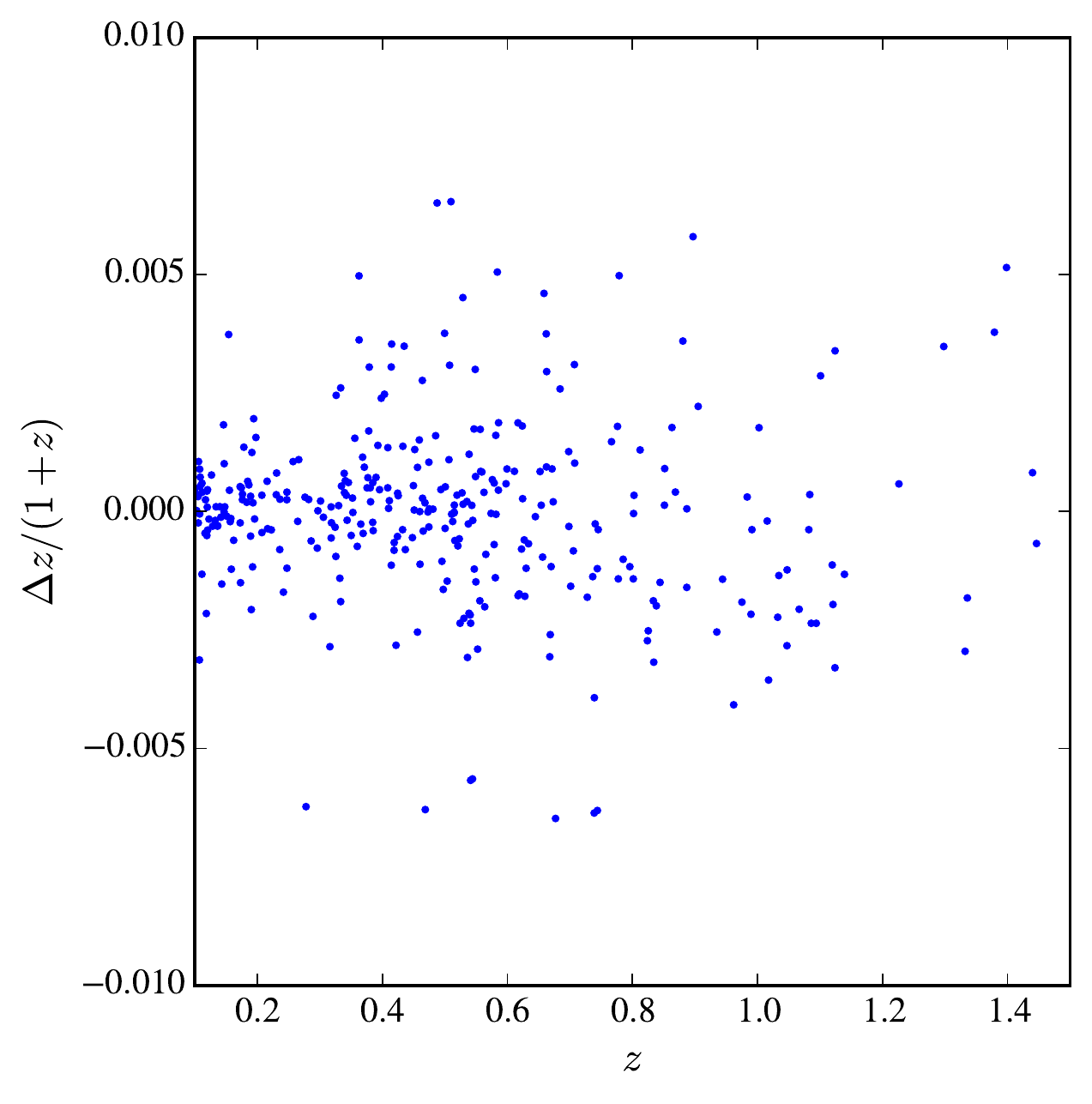}
\caption{Redshift differences between measurements from
our UVQS spectroscopy and the values listed in the
MILLIQUAS catalog.  With the exception of a few outliers
(described in the text), there is very good agreement
(RMS~$\approx 0.002$).
}
\label{fig:chkz}
\end{figure}

\section{Redshift Analysis}
\label{sec:z}

To estimate the redshift of each source, we employed modified versions
of the SDSS IDLUTILS software designed to measure quasar redshifts
in that survey \citep{sdss_qso_dr7}.  Specifically, we smoothed the 
quasar eigenspectra of SDSS (file: spEigenQSO-55732.fits)
to match the spectral resolution from each of our instruments and then
fit these eigenspectra to each spectrum, minimizing $\chi^2$.
The algorithms provide a best redshift, the model 
eigenvalues, and a statistical estimate of the redshift 
uncertainty $\sigma(z)$. 

All of the 1D spectra were visually inspected by at least two
authors using a custom GUI to assess the spectra quality. 
In parallel,  we assessed the redshift measurement by examining the
best-fit on the data.  As necessary ($\sim 30\%$ of the cases),
we performed our own estimation of the redshift by identifying
standard AGN emission features (primarily MgII and H$\beta$).
We then refitted templates to
the data using a restricted redshift interval.
We assessed the final redshift estimate based on the data quality
and the number of spectral features identified
and assigned a numerical quality assessment Z\_QUAL
with a scale of 0 (no estimate possible) 
to 5 (excellent estimate).  
Typically, sources with one prominent emission feature with a high-confidence
assignment were given 
Z\_QUAL=3.  
The majority of these are AGN with $z \approx 0.5$
where the \ion{Mg}{2} emission line occurs at $\lambda \approx 4000$\AA\
and the expected H$\beta$ emission features falls redward of our
spectral coverage.  Many of these spectra show weak Balmer emission 
(e.g.\ H$\gamma$) and/or continuum features that give high 
confidence to the reported redshift.
Furthermore, associating the detected feature to 
another emission line (e.g.\ CIII]) is strongly disfavored
due to the non-detection of other, expected features.
When multiple emission features were detected at a common
redshift, the quality of the redshift determinations 
is given 4 or 5 on our scale.
From the total candidate list 
(Tables~\ref{tab:candidates} and \ref{tab:sec}),
we measured a high-quality redshift (Z\_QUAL~$\ge 3$)
for \ngdz\ unique sources.

In the following we assume that every source with
recessional velocity $v_r \equiv zc < 500 \, \mkms$
is ``Galactic", 
which we associate to the Galaxy and members of the Local Group.
This included sources where the eigenspectra fits were poor
yet a low $v_r$ was indisputable (e.g.\ stars).
Many of these were assigned $z=0$ exactly.
The remainder of UVQS sources are assumed to be extragalactic AGN.
We caution, however, that we have not assessed
the relative line-fluxes of these sources
nor assessed the widths of emission-lines
to confirm AGN activity.  
On the other hand, every source
has a $W1-W2$ color in excess of 0.6\,mag and is therefore highly
probable to contain an AGN\footnote{The obvious exception will be
chance superpositions of two sources, which we estimate to be
a very rare occurrence ($<1\%$).}.
Furthermore, nearly all of these sources
exhibit at least one broad emission feature indicative
of an AGN.

For the redshift uncertainty of the extragalactic
sources, we have adopted the larger
of $\sigma(z)$ derived from the eigenspectra analysis
and $0.003$.  The latter value represents a systematic
uncertainty from our procedure and also allows for the 
uncertainties in deriving a systemic redshift from
broad, far-UV emission lines \citep[e.g.][]{richards02}.
We note, however, that many of the sources with $z<0.5$
exhibit [OIII] emission that may provide a smaller
redshift uncertainty.

To assess the quality of our redshift estimates, we
have compared our values against the Million Quasar
Catalog (MILLIQUAS; \mqv) compiled by \cite{milliq}.
We restrict the MILLIQUAS sample to sources with 
spectroscopic redshifts (TYPE=A or Q) and cross-matched
in RA, DEC to a 5~arcsecond radius.
In our first assessment, we noted two sources with very
large redshift difference: 
UVQSJ000856.77$-$235317.5
and UVQSJ231148.97+353541.4.  In each of our spectra,
there is a single, broad emission feature.  For 
UVQSJ000856.77$-$235317.5, we had initially identified
the feature as CIII] emission yet 
corresponding C~IV emission is not apparent.
Therefore,  we revised our evaluation to mark this
line as MgII emission and revised the redshift accordingly;
it is consistent with the previously cataloged value.
The other source is a similar case with the line identifications
reversed; we have specified the line to be MgII emission.
If the line were CIII], as previously assumed, the quasar
should have shown MgII emission.  Given that there are
also weak features at the expected wavelengths of H~$\gamma$
and H~$\beta$ for our preferred redshift, we have maintained
our estimate for the source redshift.

Figure~\ref{fig:chkz} summarizes the differences in redshifts
$\delta z \equiv \Delta z/(1+z)$ between our measurements and those previously
reported in the literature.  Ignoring the anomalous cases
described above, we measure an RMS of 0.002 for the 
191 sources with $z>0.1$.

\begin{figure}
\includegraphics[width=3.5in]{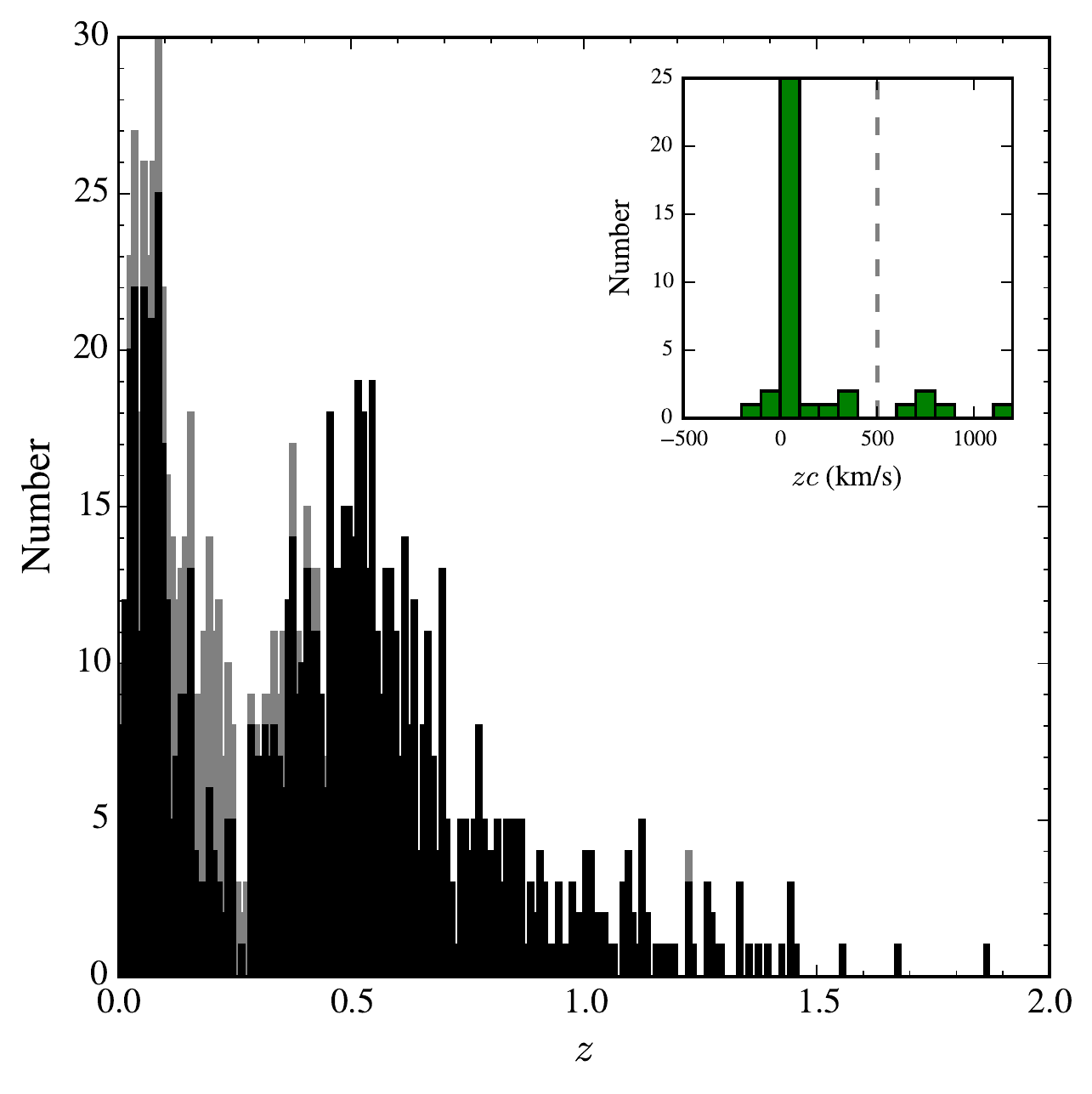}
\caption{Redshift histogram of all sources with Z\_QUAL~$>3$ from
the UVQ DR1 database.  The primary candidates (black) are
dominated by sources at $z>0.4$ with a tail to nearly 2.
In contrast, the secondary candidates (grey)
are confined to $z<0.5$ and are primarily at $z<0.2$.
These results further highlight the efficacy of our primary
$FUV-NUV$ criterion. 
The inset shows the recession velocities
$v_r \equiv z c$ of sources with $v_r \approx 0$\kms.
We associate all sources with $v_r < 500 \, \mkms$ with
the Local Group.
}
\label{fig:zhist}
\end{figure}

We present a histogram of the sources with
well-constrained redshifts (Z\_QUAL~$\ge 3$) in
Figure~\ref{fig:zhist}.  For the primary candidates (black),
there are two distributions at $z \approx 0.1$ and 
$z \approx 0.5$.  The former are low-$z$ AGN while the
other set are our desired targets.  These exhibit a 
tail of redshifts to nearly $z=2$.
As expected, the sources drawn from our
secondary list of candidates (gray) are primarily
at $z<0.3$;  only one has a redshift higher than 0.5.
Lastly, the inset to Figure~\ref{fig:zhist} shows the 
redshift measurements corresponding
to $v_r < 1000 \mkms$.  Again, we define those
with $v_r < 500 \mkms$ to be Galactic, although several
could arise from the Local Group or beyond.

\begin{figure}
\includegraphics[width=3.5in]{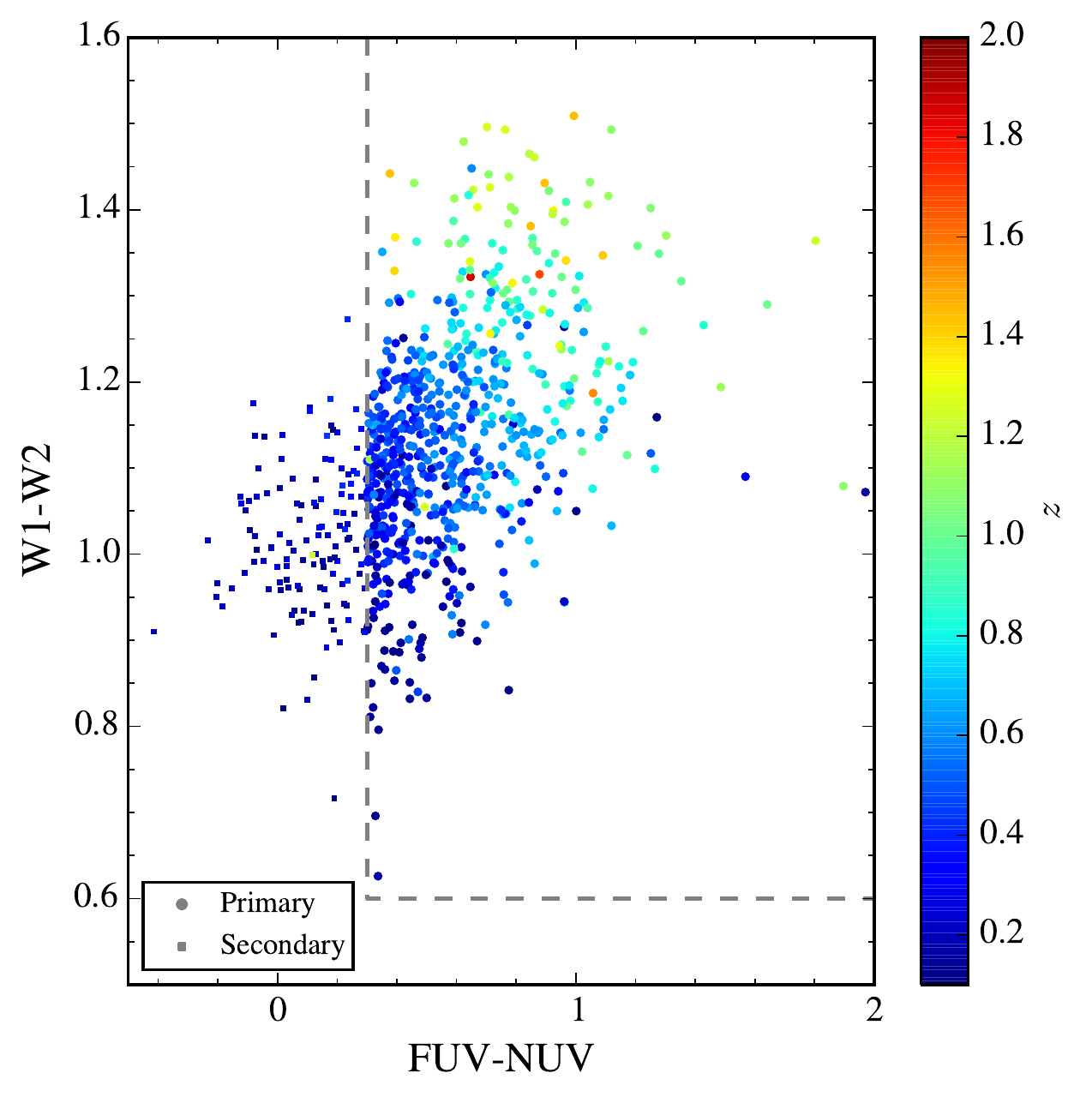}
\caption{Near-IR and UV colors of the UVQS DR1
AGN from the primary (circles) and secondary (square)
candidate lists.  The AGN show a systematic reddening
of both colors with increasing redshift.  The near-IR
evolution is related to a $k$-correction whereas
we believe the UV evolution is dominated by an increasing
average opacity to Lyman limit absorption.
}
\label{fig:dr1_colors}
\end{figure}

\section{Results}
\label{sec:results}

\begin{deluxetable}{lcccccc}
\tablewidth{0pc}
\tablecaption{UVQ DR1 AGN\label{tab:AGN}}
\tabletypesize{\scriptsize}
\tablehead{\colhead{Name} & \colhead{z} & \colhead{$\sigma(z)^a$} 
& \colhead{Z\_QUAL$^b$} & \colhead{New?$^c$}} 
\startdata 
UVQSJ000000.15-200427.7&0.291 & $0.003$&4&Y\\ 
UVQSJ000503.70-391747.9&0.652 & $0.003$&3&N\\ 
UVQSJ000609.57-261140.5&0.648 & $0.003$&3&Y\\ 
UVQSJ000741.00-635145.8&0.559 & $0.003$&3&N\\ 
UVQSJ000750.78+031733.1&1.101 & $0.003$&4&N\\ 
UVQSJ000755.67+052818.8&1.098 & $0.003$&4&Y\\ 
UVQSJ000856.77-235317.5&0.844 & $0.003$&3&N\\ 
UVQSJ001015.62-624045.1&0.850 & $0.003$&3&Y\\ 
UVQSJ001121.73-200212.1&1.226 & $0.003$&4&Y\\ 
UVQSJ001155.60-240438.8&0.767 & $0.003$&3&N\\ 
UVQSJ001521.62-385419.1&0.633 & $0.003$&3&Y\\ 
UVQSJ001637.90-054424.8&0.074 & $0.003$&5&Y\\ 
UVQSJ001641.88-312656.6&0.360 & $0.003$&5&N\\ 
UVQSJ001653.66-530932.6&0.914 & $0.003$&3&Y\\ 
UVQSJ001655.68+054822.9&1.060 & $0.003$&3&Y\\ 
UVQSJ001705.14-312536.4&0.838 & $0.003$&3&N\\ 
UVQSJ001753.32-142310.9&0.945 & $0.003$&3&Y\\ 
UVQSJ001859.75+061931.9&0.767 & $0.003$&3&Y\\ 
UVQSJ001903.85+423809.0&0.113 & $0.003$&5&Y\\ 
UVQSJ002049.31-253829.0&0.645 & $0.003$&3&N\\ 
UVQSJ002051.30-190126.8&0.962 & $0.003$&3&N\\ 
\enddata 
\tablenotetext{a}{Redshift uncertainty was derived from a template fit to the spectrum.  We report a minimum redshift error of 0.003 from systematic uncertainties.} 
\tablenotetext{b}{Redshift quality:  0=No constraint, 3=Confident, 5=Excellent} 
\tablenotetext{c}{Source is greater than 10~arcseconds offset any quasar in the MILLIQUAS catalog (\mqv) with a published spectroscopic redshift.}
\tablecomments{Table \ref{tab:AGN} is published in its entirety in the electronic edition, a portion is shown here for guidance regarding its form and content.}
\end{deluxetable}

\subsection{The UVQS Sample of New UV-Bright Quasars}

The principal goal of the UVQ Survey is to generate a new
sample of FUV-bright quasars at $z \sim 1$.
This motivated our target color criteria and subsequent
observing strategy.  With over 1000 sources analyzed,
we may reassess the survey design and efficacy.
Figure~\ref{fig:dr1_colors} presents the UV
and WISE colors of the AGN measured in UVQS~DR1, which
includes both the primary ($FUV-NUV > 0.6$\,mag)
and secondary ($-0.5 < FUV-NUV < 0.3$)
candidates.  As the source redshifts increase from
$z = 0.1$ to 2, their observed UV and near-IR 
colors redden.  We expect that the UV trend
is due primarily to Lyman limit opacity from intervening
\ion{H}{1} gas, although a flattening of the AGN SED
at approximately 1000\AA\ could 
contribute \citep[e.g.][]{telfer02,lusso+15}.
The evolution in $W1-W2$ color must be intrinsic, i.e.\
the $k$-correction for these AGN as this color shifts
from the rest-frame near-IR towards the optical
\citep[e.g.][]{assef+10,stern12}.
In hindsight, we recognize that one could more efficiently
target $z \sim 1$ quasars by adjusting the $W1-W2$
cut to a larger value (e.g.\ 1.1\,mag).

\begin{figure}
\includegraphics[width=3.5in]{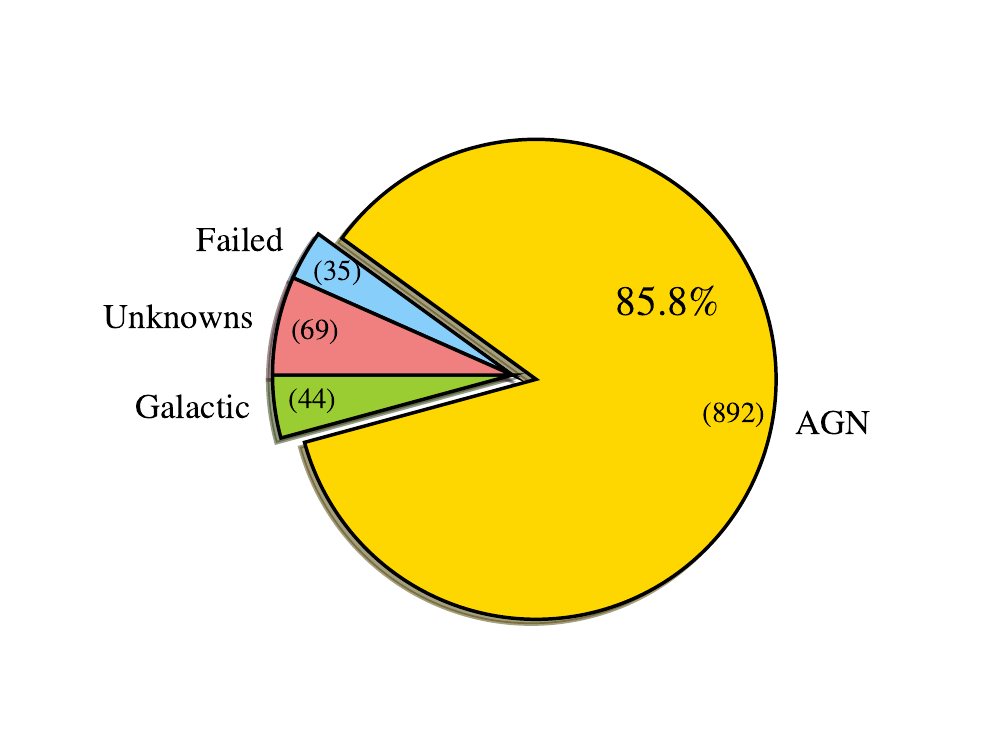}
\caption{Distribution of the source classifications for
the primary candidates observed in UVQS~DR1. 
The color-color criteria yielded a very high incidence
of AGN.  Formally, the reported rate for AGN (\agnrate)
is a lower limit as we expect many of the failed and unknown
sources are also AGN.
}
\label{fig:efficacy}
\end{figure}

\begin{figure*}
\begin{center}
\includegraphics[width=7in]{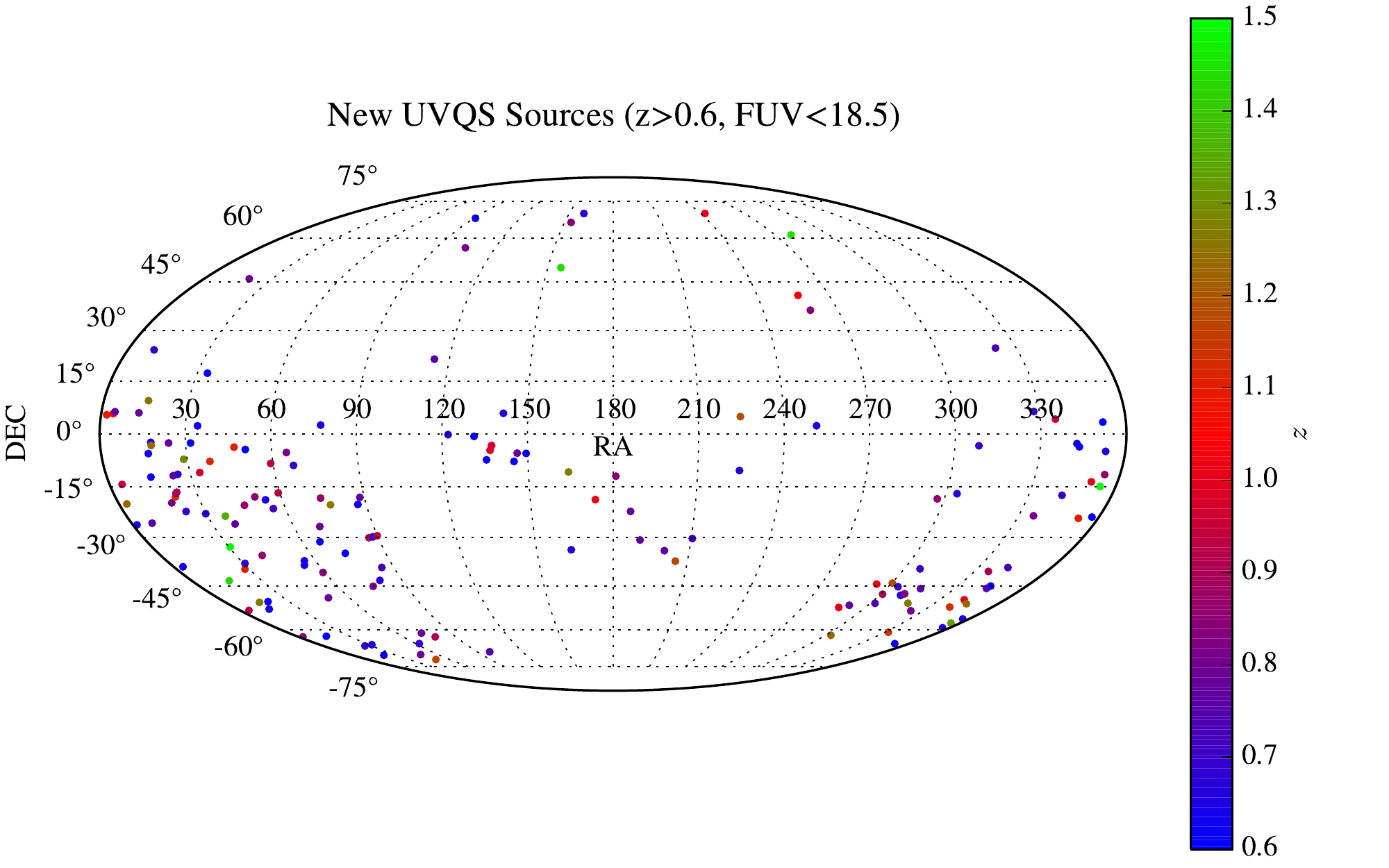}
\caption{All sky distribution of the new FUV-bright
AGN at $z>0.6$, spectroscopically confirmed in our
UVQS-DR1 survey.  The majority of these lie towards
the Southern Galactic Pole.
}
\label{fig:money}
\end{center}
\end{figure*}

The efficacy of our survey can be assessed in terms of
the fraction of AGN recovered from the total number of
sources observed.  These results are presented in 
Figure~\ref{fig:efficacy}, restricting to the primary
candidates.  Of \nprim\ primary candidates observed, 
we recovered a secure redshift for an extragalactic AGN
for \agnrate\ of the objects.  The remainder are split rather
evenly between Galactic sources, poor spectra, 
and sources without an evident spectral feature.  These
are discussed further in the following sections.

Restricting to the $z>0.6$ quasars from UVQS DR1 that
were not listed in the \mqv\ of the MILLIQUAS catalog,
Figure~\ref{fig:money} shows the sky distribution of these
new sources.  As expected, the majority of the new
discoveries occur outside of the SDSS footprint, i.e.\
towards the Southern Galactic pole.
Inspecting several of the sources within the SDSS footprint,
we find they have good photometry and expect they were simply
not targeted due to fiber collisions.

\begin{figure*}
\includegraphics[width=7in]{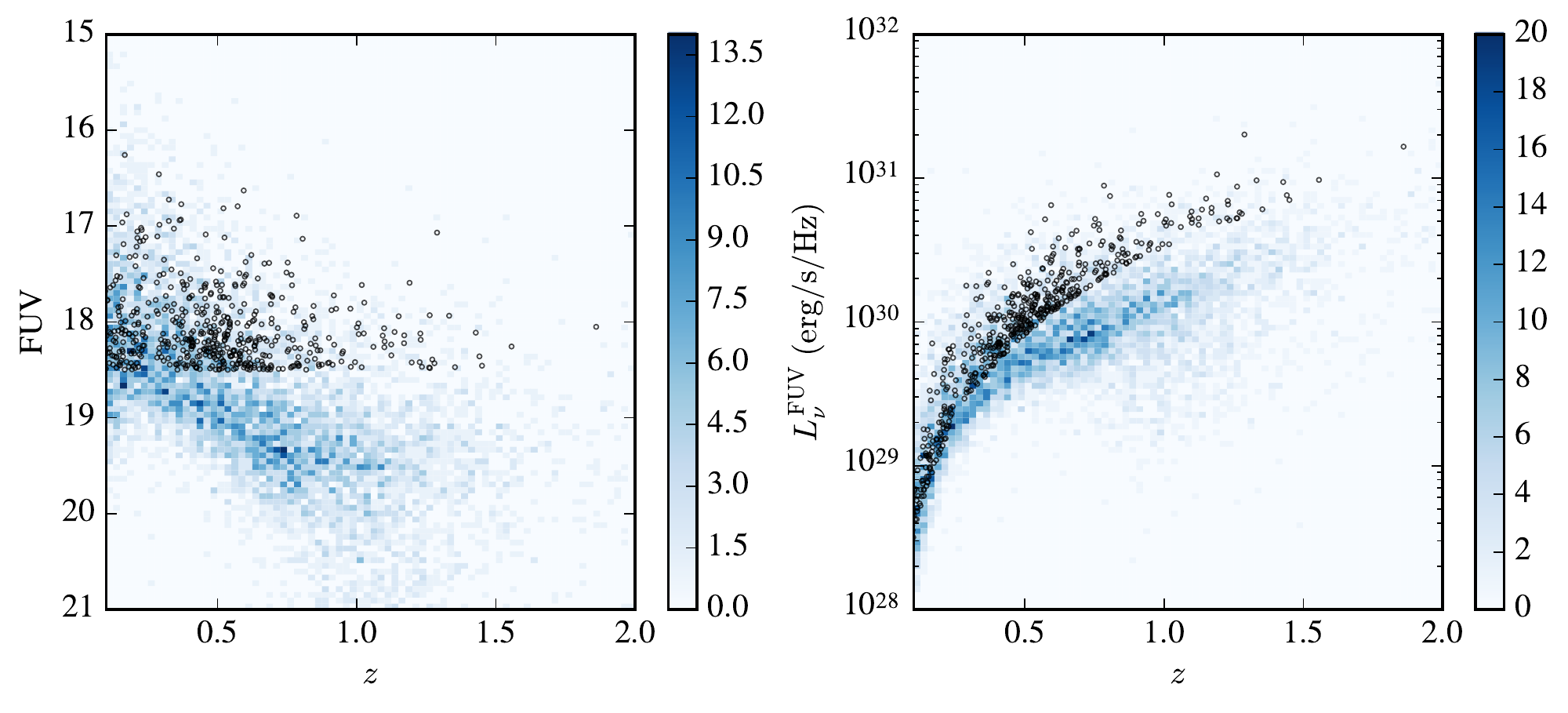}
\caption{(left) FUV GALEX magnitudes
for the AGN in the UVQS-DR1 (black dots)
compared against the locus of magnitudes from all previously
known AGN (blue, 2D histogram).  The sources
with FUV $<$ 18 mag would yield good quality
COS spectra in a modest orbit allocation.
(right) Specific FUV luminosities with the same symbol
and color coding.  At $z>0.5$, the UVQS sources represent
the most UV luminous AGN on the sky.
}
\label{fig:fuv_flux}
\end{figure*}

In Figure~\ref{fig:fuv_flux}, we compare the FUV
magnitudes and estimated luminosities (without corrections
for Galactic extinction) of the new UVQS DR1 AGN.
These are compared against previously known sources;
specifically, we show a 2D histogram of all sources
from the MILLIQUAS catalog laying within 
5~arcseconds\footnote{We caution that a small set of these previously
cataloged quasars may have erroneous redshifts 
(see $\S$~\ref{sec:z} for an example)
or are a chance coincidence match to the GALEX catalog.}
of an FUV-detected source in the GALEXGR6Plus7 photoobjall
catalog.  
At $z>0.5$, the UVQS DR1 AGN are among the brightest and
most luminous FUV sources known.  
Follow-up analysis to analyze the Eddington ratio, host galaxies,
and galactic environment of these extreme sources may
be valuable.
Given the high efficiency of our survey, we expect
that the community has now identified nearly every
FUV-bright quasar on the sky.  The only exceptions
will be within the areas not surveyed by GALEX and
the few lucky sources that shine through the dust
of the Galactic plane.

One of the most luminous quasars from our survey,
UVQSJ015454.68$-$071222.2 ($z=1.289$, $FUV=17.07$\,mag;
Figure~\ref{fig:ex_spec}),
has an interesting history worth relating.
This source was cataloged in 1962 by Haro \& Luyten
as PHL~1228 \citep{haro62}.
Based on its
color and coordinates, those authors identified the
source as a candidate
faint blue halo star towards the South Galactic pole.
Indeed, a number of their candidates have since been
confirmed as extragalactic AGN.
Clearly, a systematic redshift survey of the complete
PHL catalog is warranted.

\begin{figure*}
\includegraphics[width=7in]{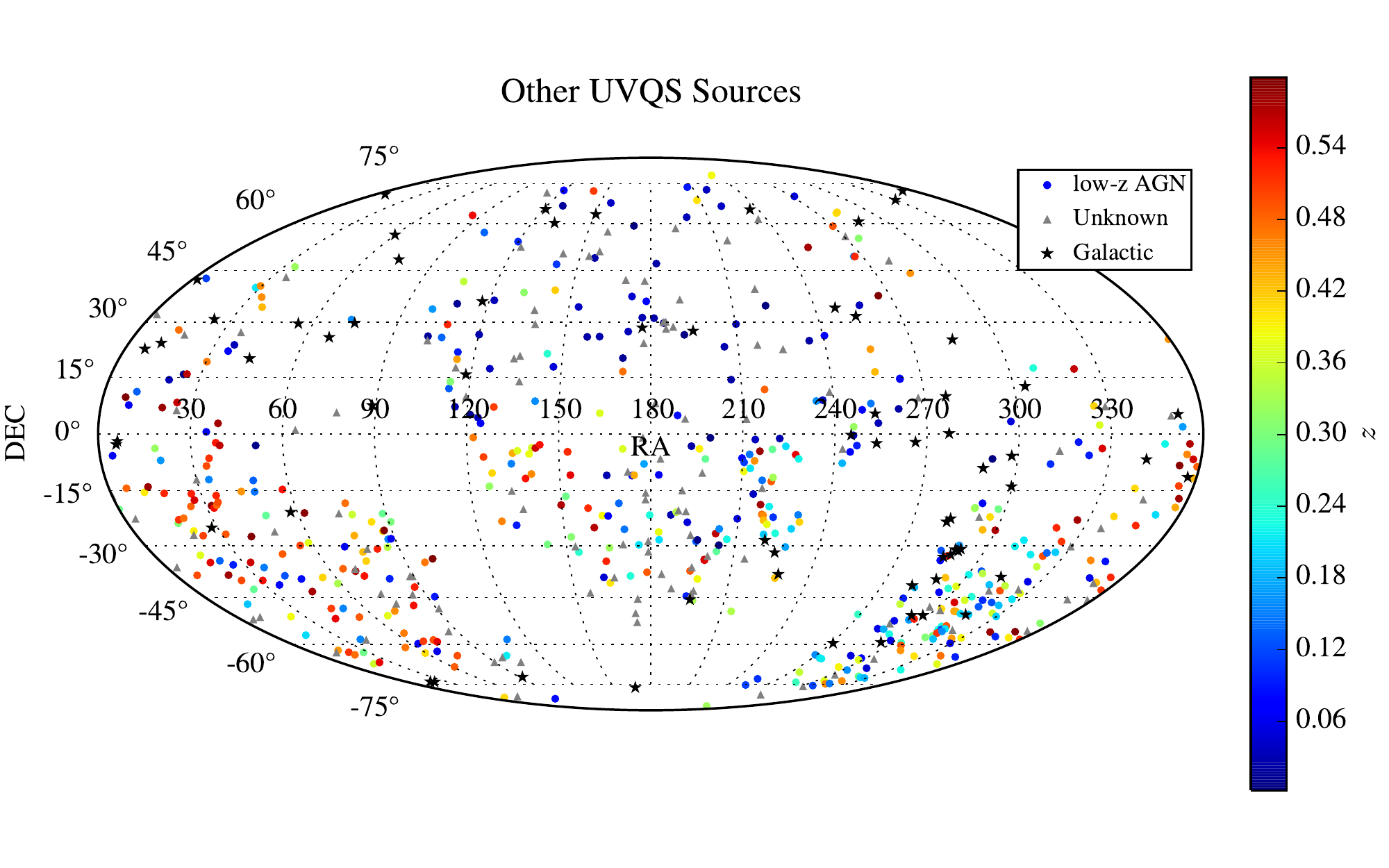}
\caption{All-sky distribution of sources other than
$z>0.6$ AGN drawn from our UVQS-DR1 dataset.
}
\label{fig:other}
\end{figure*}

\begin{deluxetable*}{lcccccc}
\tablewidth{0pc}
\tablecaption{UVQ DR1 Galactic Sources\label{tab:galactic}}
\tabletypesize{\scriptsize}
\tablehead{\colhead{Name} & \colhead{$l$} & \colhead{$b$} 
& \colhead{$W_1$} & \colhead{$W_2$} & \colhead{$E(B-V)$} \\ 
& (deg) & (deg) & (mag) & (mag) & (mag)} 
\startdata 
UVQSJ000717.69+421646.6&114.2718 & $-19.8486$&12.44 & 11.51&0.07\\ 
UVQSJ002255.11-024418.7&106.0850 & $-64.6733$&13.25 & 12.12&0.03\\ 
UVQSJ002324.11+704009.9&120.5946 & $7.9250$&7.28 & 6.58&0.95\\ 
UVQSJ002452.54-015335.4&107.6594 & $-63.9745$&9.56 & 8.69&0.03\\ 
UVQSJ002715.37+224158.1&115.6634 & $-39.8307$&13.15 & 11.96&0.04\\ 
UVQSJ004433.61+241919.7&120.9291 & $-38.5229$&11.41 & 10.81&0.05\\ 
UVQSJ011219.70-735126.0&300.9427 & $-43.1902$&9.66 & 8.52&0.04\\ 
UVQSJ012138.72-735841.0&300.0898 & $-42.9831$&9.98 & 9.31&0.05\\ 
UVQSJ013450.10+305445.0&133.7961 & $-31.0421$&14.86 & 13.79&0.05\\ 
UVQSJ015159.68-250314.9&207.6540 & $-76.2551$&17.82 & 16.34&0.01\\ 
UVQSJ025637.57+200537.2&158.9238 & $-33.8856$&7.84 & 7.22&1.24\\ 
UVQSJ033900.56+294145.6&161.1830 & $-20.4629$&7.61 & 6.83&0.22\\ 
UVQSJ035056.00-204815.9&214.1511 & $-48.7234$&9.66 & 9.02&0.07\\ 
\enddata 
\tablecomments{UVQS DR1 sources with recessional velocity $v_r < 500 \mkms$. Reddening $E(B-V)$ estimates are based on the \cite{sfd98} extinction maps.} 
\tablecomments{Table \ref{tab:galactic} is published in its entirety in the electronic edition, a portion is shown here for guidance regarding its form and content.}
\end{deluxetable*} 

\begin{deluxetable}{lccccc}
\tablewidth{0pc}
\tablecaption{UVQ DR1 Unknown Sources\label{tab:unknowns}}
\tabletypesize{\scriptsize}
\tablehead{\colhead{Name} & \colhead{$l$} & \colhead{$b$} 
& \colhead{$FUV$} & \colhead{$NUV$} \\ 
& (deg) & (deg) & (mag) & (mag)} 
\startdata 
UVQSJ000009.65-163441.4&71.9317 & $-74.1194$&18.48 & 17.72\\ 
UVQSJ001444.02-223522.6&59.5364 & $-80.5220$&18.39 & 17.34\\ 
UVQSJ001529.53-360535.3&341.1397 & $-78.2250$&18.23 & 17.70\\ 
UVQSJ004038.09-505756.5&307.1282 & $-66.0744$&17.43 & 16.77\\ 
UVQSJ005116.64-624204.3&302.9636 & $-54.4270$&18.31 & 17.84\\ 
UVQSJ010018.69-741815.9&302.1140 & $-42.8097$&17.55 & 17.12\\ 
UVQSJ012031.66-270124.6&213.6632 & $-83.5246$&18.20 & 17.36\\ 
UVQSJ013955.76+061922.4&144.0255 & $-54.5508$&17.64 & 17.19\\ 
UVQSJ022239.60+430207.8&140.1429 & $-16.7669$&17.70 & 16.88\\

\enddata 
\tablecomments{UVQS DR1 sources with good spectral quality but where no precise redshift could be measured.} 
\tablecomments{Table \ref{tab:unknowns} is published in its entirety in the electronic edition, a portion is shown here for guidance regarding its form and content.}
\end{deluxetable}

\begin{figure}
\includegraphics[width=3.5in]{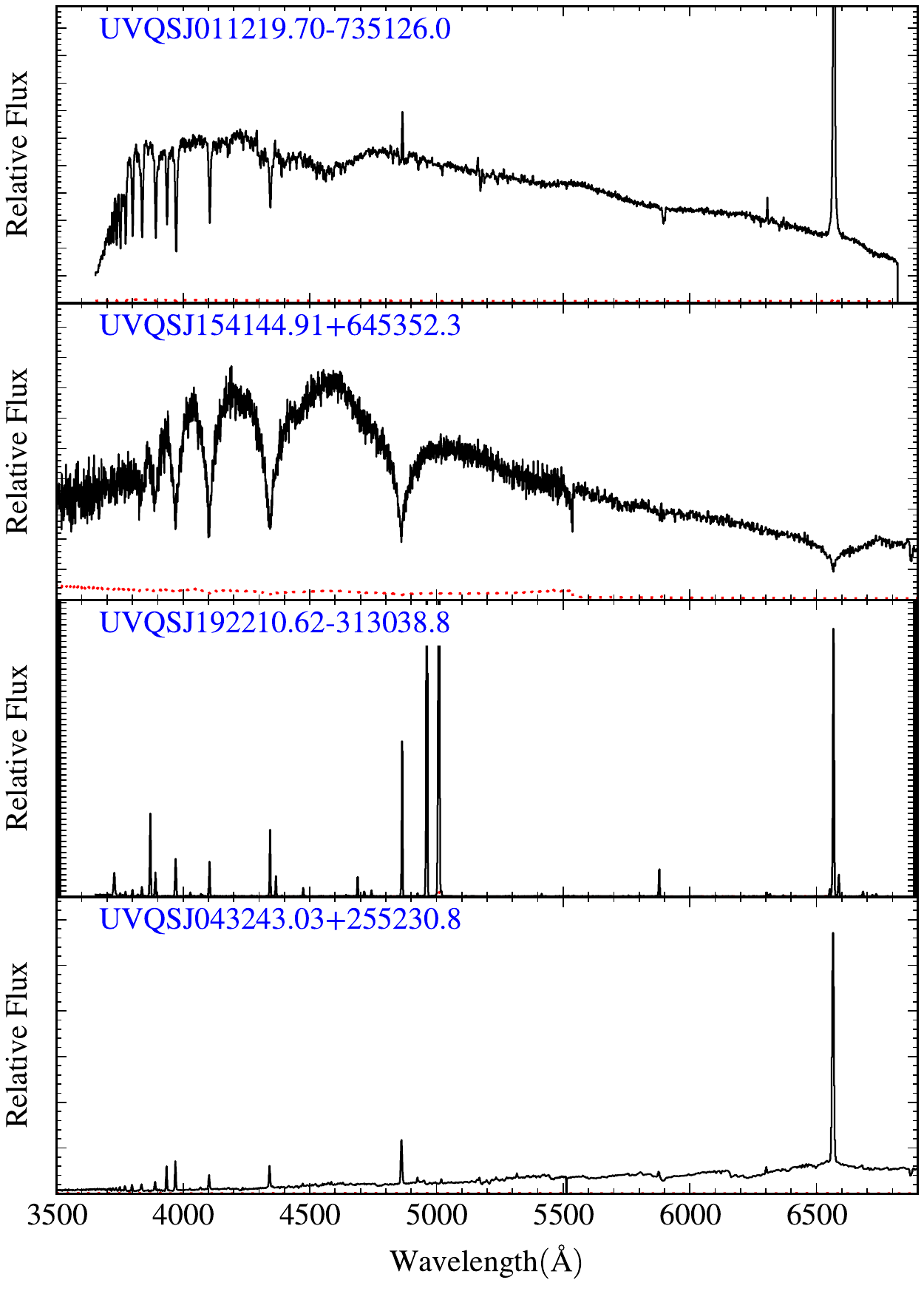}
\caption{UVQS-DR1 spectra for a representative set
of Galactic sources unintentionally observed in our survey.
}
\label{fig:galactic}
\end{figure}

\subsection{Other Sources}

Figure~\ref{fig:other} shows an all-sky plot of the 
other UVQS sources: 
AGN at $z<0.6$,
sources with good spectra but without a precise 
redshift, and Galactic sources.
Not surprisingly, the latter are primarily located
near the Galactic plane.
In DR1, we observed \ngal\ sources satisfying our
color criteria (including 24 with $FUV-NUV < 0.6$\,mag)
whose spectra yield a recessional velocity $v_r < 500 \mkms$.
These are listed in Table~\ref{tab:galactic}.
Spectra for a representative set is shown
in Figure~\ref{fig:galactic}.
These objects include hot stars, white dwarfs, 
planetary nebulae, and Herbig Ae/Be stars, all of which
have high surface temperatures explaining
their high UV fluxes.
It is more difficult, however, to explain their
$W1-W2$ color.  Several of the sources have {\it WISE}
fluxes near their detection limit, i.e.\ poor photometry
may explain their inclusion.  Another set have substantial
extinction ($E(B-V)>0.3$\,mag).
The remainder, however, may be chance super-positions
with a low mass star.
Finally, we note that
from the full set of Galactic sources we identify
a small sample with highly unusual spectra 
(e.g.\ Margon et al.\ 2015, submitted).

There are 93 sources with a good quality
spectrum (SPEC\_QUAL~$\ge 3$) for which we cannot
recover a secure redshift.
The majority of these have been previously cataloged
as blazars (or BL Lac objects).  Examining Figure~\ref{fig:other}
we note these sources are distributed across the sky,
consistent with an extragalactic origin.
Table~\ref{tab:unknowns} lists the sample of these
unknowns.

Finally, \nmiss\ of the brightest primary candidates
($FUV < 17.5$\,mag) went unobserved.  Nearly all of these
are well resolved in the SDSS or DSS imaging and were
dismissed as having $z \ll 1$.
Three of the sources --
J124735.07-035008.2,
J221153.89+184149.9,
J221712.27+141420.9 
--
went unobserved due to errors in book-keeping 
or insufficient observing time.
We will endeavor to provide spectra of these sources in our
second data release.

\section{Concluding Remarks}

We have performed an all-sky survey for $z \sim 1$, 
FUV-bright quasars selected from GALEX and WISE photometry.
The majority of these candidates lay towards the Southern
Galactic Pole, i.e.\ outside the SDSS footprint.
We confirmed 256 AGN at $z>0.6$, 155 of which had no previously
reported spectroscopic redshift.  
Altogether, the UVQS DR1 includes \newagn~previously
uncataloged AGN with FUV~$< 18$\,mag which are excellent targets for
absorption-line analysis using {\it HST}/COS.
Indeed, a handful of these AGN are already scheduled for
Cycle~24 observations.
In our second data release of UVQS, we expand the
search to NUV-bright AGN at $z \sim 1$.

\clearpage

\acknowledgements

We kindly thank Kate Rubin and Neil Crighton for their
twilight observations on several candidates.  T. R. M. and J. T. acknowledge support for this project from the STScI Director's Discretionary Research Fund under allocation D0001.82451. 
J. X. P. and N. T. acknowledge partial support from the National
Science Foundation (NSF) grant AST-1412981. 
JFH acknowledges generous support from the Alexander von Humboldt
foundation in the context of the Sofja Kovalevskaja Award. The
Humboldt foundation is funded by the German Federal Ministry for
Education and Research.  

This work is based on data obtained from Lick Observatory, owned and operated by the University of California.  We thank the Mount Hamilton staff of Lick Observatory for assistance in acquiring the observations.

Based on observations collected at the Centro Astron\'omico 
Hispano Alem\'an (CAHA) at Calar Alto, operated jointly by the Max-Planck Institut 
fur Astronomie and the Instituto de Astrof\'isica de Andaluc\'ia (CSIC).

Some of the data presented herein were obtained at the W.M. Keck
Observatory, which is operated as a scientific partnership among the
California Institute of Technology, the University of California, and
the National Aeronautics and Space Administration. The Observatory was
made possible by the generous financial support of the W.M. Keck
Foundation.  Some of the Keck data were obtained through the NSF
Telescope System Instrumentation Program (TSIP), supported by AURA
through the NSF under AURA Cooperative Agreement AST 01-32798 as
amended. 
The authors wish to recognize and acknowledge the very
significant cultural role and reverence that the summit of Mauna Kea
has always had within the indigenous Hawaiian community. We are most
fortunate to have the opportunity to conduct observations from this
mountain.

This publication makes use of data products from the Wide-field Infrared Survey Explorer, which is a joint project of the University of California, Los Angeles, and the Jet Propulsion Laboratory/California Institute of Technology, and NEOWISE, which is a project of the Jet Propulsion Laboratory/California Institute of Technology. WISE and NEOWISE are funded by the National Aeronautics and Space Administration.




\end{document}